\documentclass[twocolumn,showpacs,prl,aps]{revtex4}
\usepackage{amssymb}
\usepackage{amsmath}
\usepackage{amsfonts}
\usepackage{graphics}
\usepackage{epic}
\usepackage{eepic}
\usepackage{color}
\usepackage{epsfig}

\begin{document}

\def\ra{\rangle}
\def\la{\langle}
\def\bege{\begin{equation}}
\def\ende{\end{equation}}
\def\begarr{\begin{eqnarray}}
\def\endarr{\end{eqnarray}}
\def\ha{{\hat a}}
\def\hb{{\hat b}}
\def\hu{{\hat u}}
\def\hv{{\hat v}}
\def\hc{{\hat c}}
\def\hd{{\hat d}}
\def\no{\noindent}\def\non{\nonumber}
\def\hi{\hangindent=45pt}
\def\v{\vskip 12pt}

\newcommand{\bra}[1]{\left\langle #1 \right\vert}
\newcommand{\ket}[1]{\left\vert #1 \right\rangle}

\title{Optical Zeno Gate: Bounds for Fault Tolerant Operation}

\author{Patrick M. Leung} \email{pmleung@physics.uq.edu.au}
\author{Timothy C. Ralph} %\email{ralph@physics.uq.edu.au}

\affiliation{Centre for Quantum Computer Technology, Department of
Physics,  University of Queensland, Brisbane 4072, Australia}

\date{\today}
%\date{March 31, 2002}

\begin{abstract}

In principle the Zeno effect controlled-sign gate of Franson et al's
(PRA 70, 062302, 2004) is a deterministic two-qubit optical gate.
However, when realistic values of photon loss are considered its
fidelity is significantly reduced. Here we consider the use of
measurement based quantum processing techniques to enhance the
operation of the Zeno gate. With the help of quantum teleportation,
we show that it is possible to achieve a Zeno CNOT gate (GC-Zeno
gate) that gives (near) unit fidelity and moderate probability of
success of 0.76 with a one-photon to two-photon transmission ratio
$\kappa=10^4$. We include some mode-mismatch effects and estimate
the bounds on the mode overlap and $\kappa$ for which fault tolerant
operation would be possible.

%We have modelled the Zeno effect Control-Sign gate of Franson et al
%(PRA 70, 062302, 2004) and shown that high two-photon to one-photon
%absorption ratios, $\kappa$, are needed for high fidelity free
%standing operation. Hence we instead employ this gate for cluster
%state fusion, where the requirement for $\kappa$ is less
%restrictive. With the help of partially offline one-photon and
%two-photon distillations, we can achieve a fusion gate with unity
%fidelity but non-unit probability of success. We conclude that for
%$\kappa > 2200$, the Zeno fusion gate will out perform the
%equivalent linear optics gate.
\end{abstract}

\maketitle

%\begin{equation}{2}

\section{I. Introduction}

Photons are a natural choice for making qubits because the quantum
information encoded can have a long decoherence time and is easy to
manipulate and measure. Also, photonic qubits are the only type of
qubits that are feasible for long distance quantum communication.
Quantum information processing requires universal two-qubit
entangling gates. Knill et al~\cite{Knill01} showed that it is
theoretically possible to do scalable quantum computing with linear
optics by using measurement induced interactions to perform the
two-qubit gates. However, despite a continuous effort in reducing
the resource requirement~\cite{Yoran03,Nielsen04,Hayes04,Brown05},
the resource overhead is still high for linear optical quantum
computing. Franson et al\cite{Franson04} proposed the use of
two-photon absorption non-linearity and exploiting the quantum Zeno
effect to implement a control sign (CZ) gate that requires much less
resources than linear optics schemes. However, the problem with the
Zeno gate is that photon loss affects the performance of the Zeno
gate significantly and the single photon to two-photon loss ratio
requirement is very stringent. One solution couldbe to combine
measurement induced quantum processing with the Zeno gate.
Previously we have shown that when using the Zeno gate for qubit
fusion~\cite{Leung06}, state distillation~\cite{Thew01} with
post-selection can boost the gate fidelity to unity and that for
less stringent absorption ratios the gate has an advantage in
success probability over linear optics gates for fusing clusters of
qubits. These clusters of qubits can then be used for cluster state
quantum computing~\cite{Raussendorf01}. In addition, Myers and
Gilchrist~\cite{Myers06} have shown that the performance of the Zeno
gate may be enhanced by using error correction codes such as
redundancy and parity encoding.

Here we design a high fidelity Zeno CNOT gate suitable for
circuit-based quantum computing. Although with the current estimate
of the photon loss ratio, only a poor fidelity Zeno gate is directly
achievable, here we show that it is possible to use two of these
Zeno gates to do Bell measurements and implement a
Gottesman-Chuang\cite{Gottesman99} teleportation type of CNOT gate
(GC-Zeno gate) that, like the fusion gate, gives high fidelity via
state distillation and moderate success probability via partially
off-line state preparation. We include the effect of mode-mismatch
and detector efficiency on the scheme and estimate lower bounds on
the parameter which in principle allow fault tolerant operation.\\

The paper is arranged in the following way. The introduction
continues with a subsection on the Zeno CZ gate, which describes the
scheme and modelling of the gate and give descriptions on the
modelling parameters that are also used for modelling the GC-Zeno CZ
gate. In section II, we discuss the GC-Zeno gate and the effect of
mode-mismatch and detector efficiency on the gate. In section III,
we give estimates of the lower bounds on the photon loss ratio and
mode-matching, followed by a subsection on the advantage in using
state distillation. We conclude and summarize in section IV.

\subsection{Ia. Zeno CZ Gate}

Franson et al's control sign gate scheme consists of a pair of
optical fibers weakly evanescently coupled and doped with two-photon
absorbing atoms. The purpose of the two-photon absorbers is to
suppress the occurrence of two photon state components in the two
fibre modes via the Quantum Zeno effect. This allows the state to
remain in the computational basis. After a length of fibre
corresponding to a complete swap of the two fibre modes, a $\pi$
phase difference is produced between the $|11 \rangle$ term and the
other three basis terms. After swapping the fibre modes by simply
crossing them, a CZ gate is achieved. The gate becomes near
deterministic and performs a near unitary operation when the Quantum
Zeno effect is strong and photon loss is insignificant. However,
with current technology, the strength of the Quantum Zeno effect is
a few orders of magnitude below what is required, and thus the Zeno
gate has significant photon loss.\\

\begin{figure}[h]
\centerline{\psfig{figure=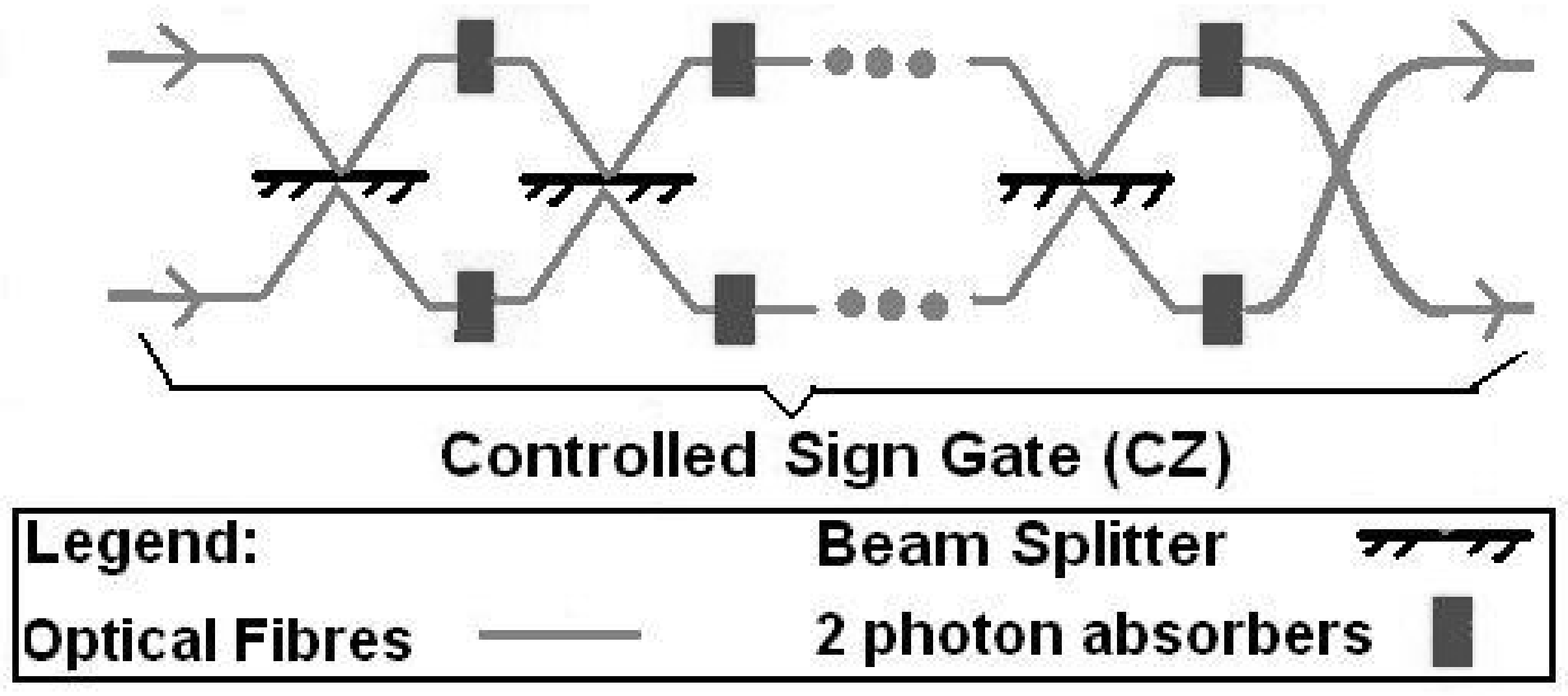,width=2.5cm,angle=90}}
\caption{Construction of our CZ gate.\\} \label{fig:OurCsign}
\end{figure}

\begin{figure}[h]
\centerline{\psfig{figure=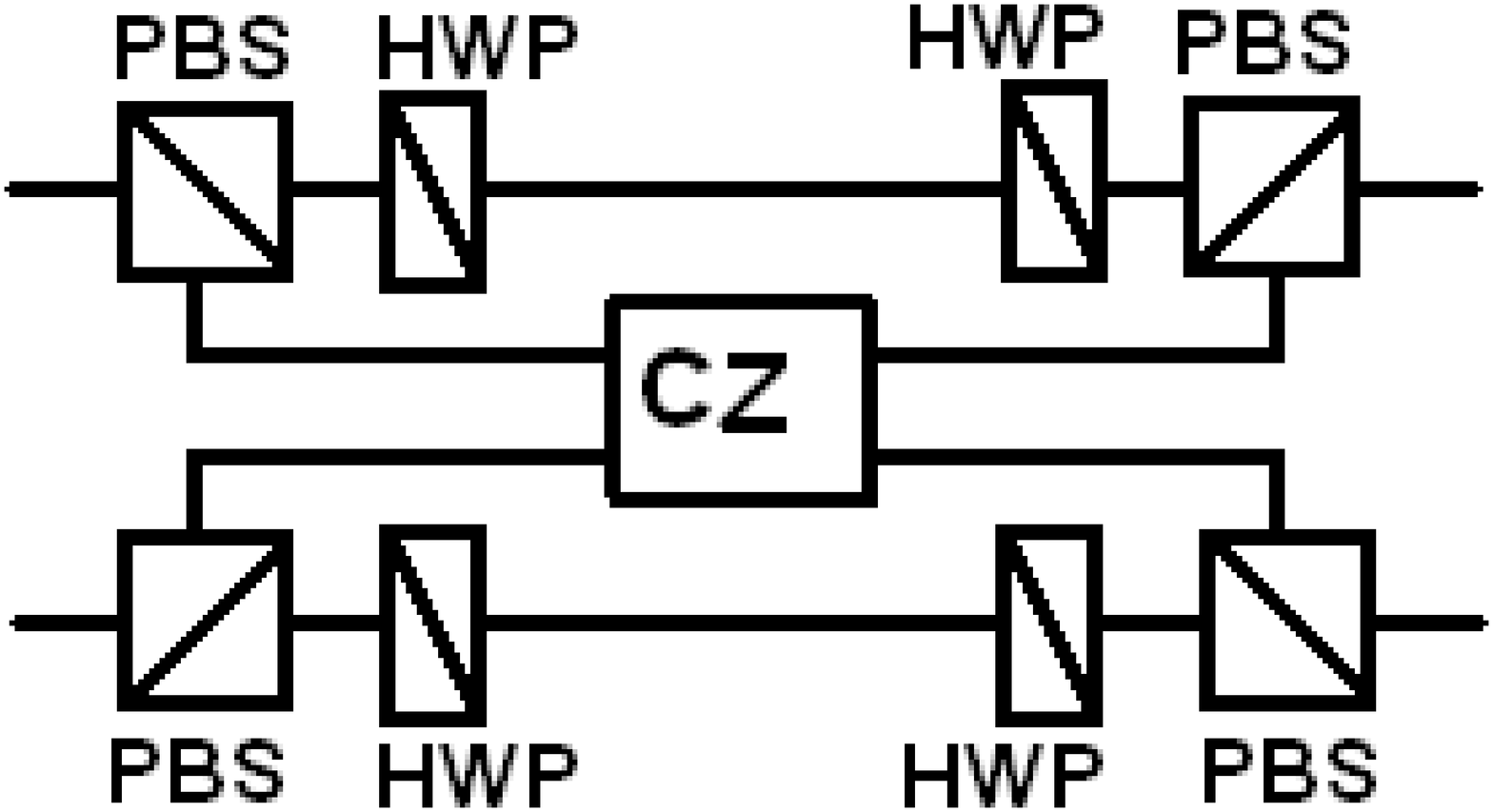,width=2cm,angle=90}} \caption{CZ
gate in dual rail implementation.\\}\label{fig:CZ}
\end{figure}

In~\cite{Leung06}, the gate is modelled as a succession of $n$ weak
beamsplitters followed by two-photon absorbers as shown in
Fig.~\ref{fig:OurCsign}. As $n \to \infty$ the model tends to the
continuous coupling limit envisaged for the physical realization.
The gate operates on the single-rail encoding for which
$|0\ra_{L}=|0\ra$ and $|1\ra_{L}=|1\ra$ with the kets representing
photon Fock states. Fig.~\ref{fig:CZ} shows how the single rail CZ
can be converted into a dual rail~\cite{Lund02} CZ with logical
encoding $|0\ra_{L}=|H\ra$ and $|1\ra_{L}=|V\ra$. Let $L$ be the
total length of $n$ absorbers. Also, let
$\gamma_{1}=\exp(\frac{-\lambda}{n\kappa})$ and
$\gamma_{2}=\exp(\frac{-\lambda}{n})$ be the probability of single
photon and two-photon transmission respectively for one absorber.
Here the parameter $\lambda=\chi L$, where $L$ is the length of the
absorber and $\chi$ is the corresponding proportionality constant
related to the absorption cross section. Furthermore, $\kappa$
specifies the relative strength of the two transmissions and relates
them by $\gamma_{2}=\gamma_{1}^{\kappa}$. This CZ gate does the
following operation:

\begin{eqnarray}
|00\ra & \rightarrow & |00\ra\nonumber\\
|01\ra & \rightarrow & \gamma_{1}^{n/2}|01\ra\nonumber\\
|10\ra & \rightarrow & \gamma_{1}^{n/2}|10\ra\nonumber\\
|11\ra & \rightarrow & -\gamma_{1}^{n}\tau|11\ra + f(|02 \rangle,
|20 \rangle) \label{eqn:incomplete}
\end{eqnarray}
where the new expression for $\tau$ is given by:
\begin{eqnarray}
\tau_{n,\lambda} & = & \frac{2^{-\frac{3}{2}-n}}{d_{n,\lambda}}\bigl( (g_{n,\lambda}+\frac{d_{n,\lambda}}{\sqrt{2}})^{n}(\sqrt{2}d_{n,\lambda}-h_{n,\lambda}) \nonumber\\
& & + (g_{n,\lambda}-\frac{d_{n,\lambda}}{\sqrt{2}})^{n}(\sqrt{2}d_{n,\lambda}+h_{n,\lambda})\bigr)\nonumber\\
d_{n,\lambda} & = & \sqrt{(1+\cos\frac{2\pi}{n})(1+\gamma_{2})+2\sqrt{\gamma_{2}}(\cos(\frac{2\pi}{n})-3)}\nonumber\\
g_{n,\lambda} & = & (\cos\frac{\pi}{n})(\sqrt{\gamma_{2}}+1)\nonumber\\
h_{n,\lambda} & = & 2(\cos\frac{\pi}{n})(\sqrt{\gamma_{2}}-1)
\end{eqnarray}

The explicit form of the $|02 \rangle, |20 \rangle$ state components
are suppressed, as they lie outside the computational
basis and so do not explicitly contribute to the fidelity.\\

\begin{figure}[h]
\centerline{\psfig{figure=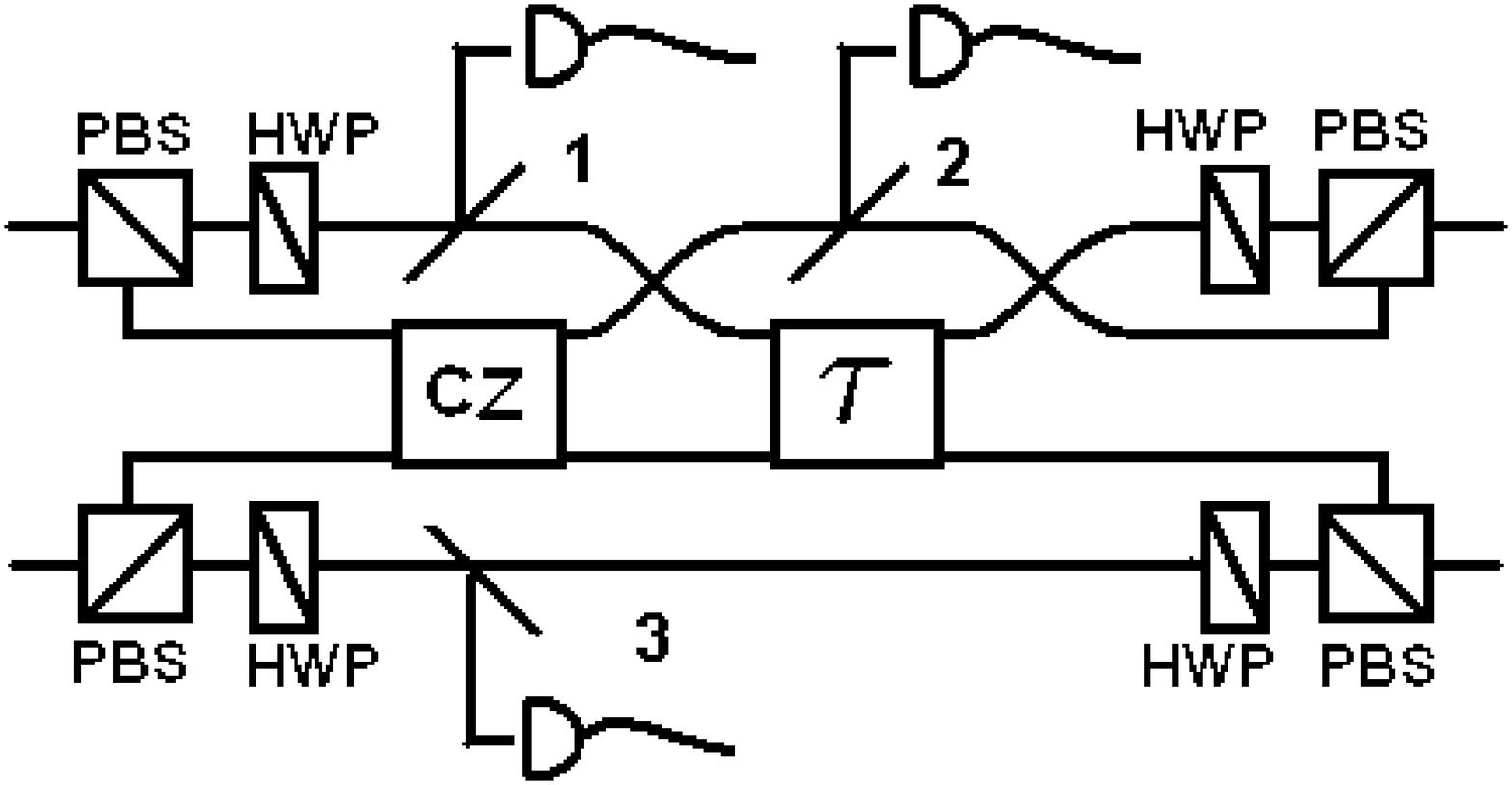,width=2cm,angle=90}}
\caption{Schematic of distillated Zeno CZ gate.\\}
\label{fig:two_photon_distill}
\end{figure}

From equation~\ref{eqn:incomplete}, it is clear that the amplitude
of the four computational states are unequal and this lowers the
gate fidelity. With the current best estimate of $\kappa=10^4$, the
unherald fidelity is only 0.94. If the gate is used in a measurement
based strategy then state distillation can be used and the fidelity
of the gate can be improved by trading off some success probability.
Figure~\ref{fig:two_photon_distill} shows the distillated Zeno CZ
gate circuit. The $\tau$ gate is simply an interferometer consisting
of two 50-50 beam splitters with a two-photon absorber in each arm,
which gives operation: $|00\ra \rightarrow |00\ra$, $|01\ra
\rightarrow \sqrt{\gamma_{1}'}|01\ra$, $|10\ra \rightarrow
\sqrt{\gamma_{1}'}|10\ra$, and $|11\ra \rightarrow
\gamma_{1}'\tau|11\ra$. Here $\gamma_1^{'}=\tau^{1/\kappa}$ is the
single photon transmission coefficient of the absorber. The
distillation beam splitters labelled 1 to 3 have transmission
coefficient $\sqrt{\gamma_1^n}$, $\sqrt{\gamma_1^{'}}$,
$\sqrt{\gamma_1^{n}\gamma_1^{'}}\tau$ respectively. With these
distillations in place, the operation of the distillated Zeno CZ
gate is:

\begin{eqnarray}
|00\ra & \rightarrow & \gamma_{1}^{n}\gamma_{1}^{'}\tau|00\ra\nonumber\\
|01\ra & \rightarrow & \gamma_{1}^{n}\gamma_{1}^{'}\tau|01\ra\nonumber\\
|10\ra & \rightarrow & \gamma_{1}^{n}\gamma_{1}^{'}\tau|10\ra\nonumber\\
|11\ra & \rightarrow & -\gamma_{1}^{n}\gamma_{1}^{'}\tau|11\ra +
f(|02 \rangle, |20 \rangle) \label{eqn:distillated_CZ}
\end{eqnarray}

After measuring the output (detectors at output are not shown in
fig.\ref{fig:two_photon_distill}) and treating the photon bunching
terms ($|02\ra$, $|20\ra$) and the terms with photon loss as
failures (excluded from Eq.~\ref{eqn:distillated_CZ} for clarity),
renormalising the states gives unit heralded fidelity independent of
$\lambda$ and probability of success $P_s =
\gamma^{2n}_{1}\gamma_{1}'^{2}\tau^{2} =
e^{-2\lambda/\kappa}\tau^{2+2/\kappa}$.

\section{II. Zeno Gate Using Gottesman-Chuang Scheme}

As argued above, state distillation can improve the fidelity of the
Zeno gate to unity by trading off success probability. However, the
output of the distillated Zeno gate contains terms outside the
computational basis due to photon loss and photon bunching. Hence if
we want the gate to have unit fidelity, it is necessary to measure
the output and exclude these failure terms by post-selection.
However, such post-selection means that the gate can no longer be
used directly as a CNOT gate for circuit-based quantum computing.\\

\begin{figure}
\centerline{\psfig{figure=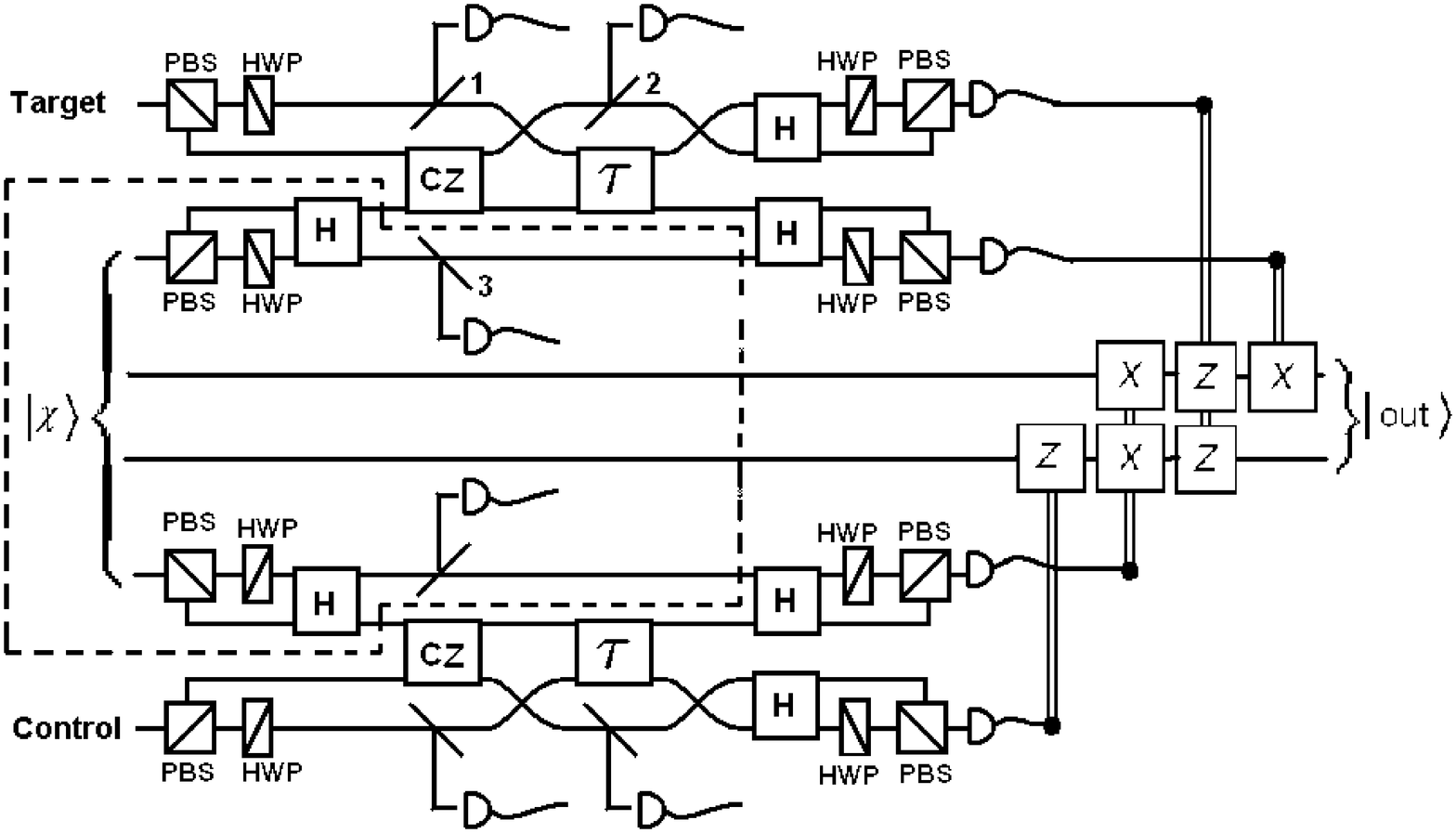,width=3.5cm,angle=90}}
\caption{Schematic of GC-Zeno gate. The state $|\chi\ra$ is
$((|00\ra+|11\ra)|00\ra+(|01\ra+|10\ra)|11\ra)/2$\\}
\label{fig:GC_Zeno}
\end{figure}

\begin{figure}
\centerline{\psfig{figure=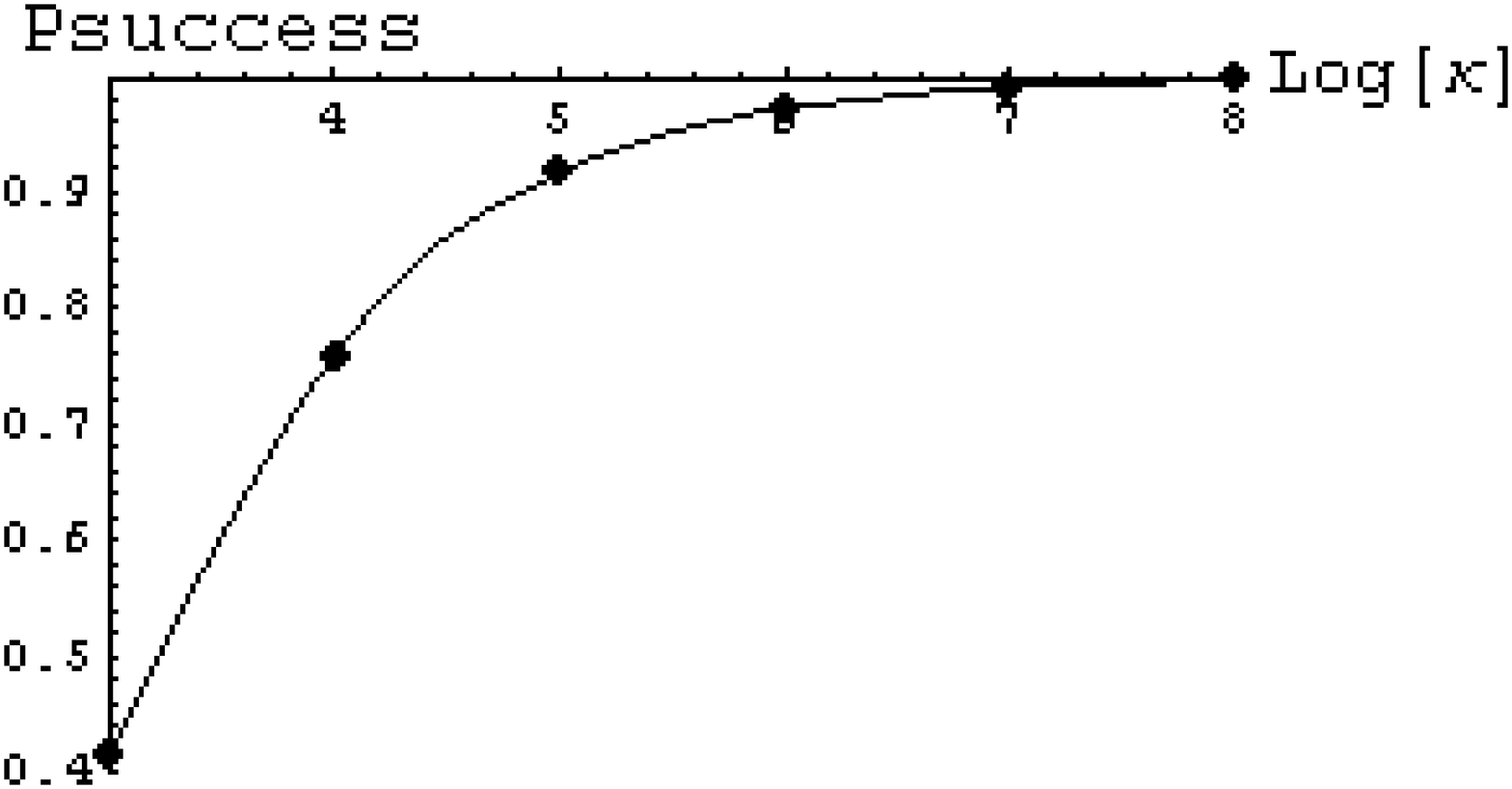,width=2.5cm,angle=90}}
\caption{Plot of probability of success versus $\log(\kappa)$ (in
base 10) for GC-Zeno gate. Note that the success probability is not
one at $\kappa=10^8$, but that the curve asymptotically approaches
one for very large $\kappa$. Result is per two input qubits.
Detector inefficiency is taken into account in accord with Dawson et
al's bound.} \label{fig:Psuccess_GC_Zeno}
\end{figure}

Gottesman and Chuang~\cite{Gottesman99} showed the viability in
using state teleportation and single qubit operations to implement a
CNOT gate. The scheme requires the four qubit entangled state
$|\chi\ra=((|00\ra+|11\ra)|00\ra+(|01\ra+|10\ra)|11\ra)/2$.
Preparing the entangled state requires a CZ operation, which can be
done off-line with linear optics with high fidelity. Bell
measurements are made between the input qubits and the first and
last qubits of $|\chi\ra$. The measurement results are fed forward
for some single qubit corrections such that the circuit gives a
proper CNOT operation. Here we propose using such scheme, as shown
in figure~\ref{fig:GC_Zeno}, to implement a GC-Zeno CNOT gate with
high fidelity.
%By having one
%hadamard operation on the entangled-resource qubit at input and
%another hadamard operation on the each of the output qubits after
%the distillated Zeno CZ gate, we can perform a Bell measurement (see
%fig.~\ref{fig:GC_Zeno}). With two of these Bell measurements, each
%for both input qubits, plus some single qubit operations, a
%teleportation type of Zeno CNOT gate is obtained.
Since this gate includes state distillation, post-selection and
off-line state preparation, the gate has unit fidelity (under
perfect mode-matching) and moderate success probability.
Figure~\ref{fig:Psuccess_GC_Zeno} plots the probability of success
against the one-photon to two-photon transmission ratio $\kappa$. It
shows that with $\kappa=10^4$ (current best estimate), the
probability of success is about 0.76, which is better than the break
even point of 0.25 for the linear optics version of this
gate~\cite{Knill01}.\\

\subsection{IIa. Effect of Mode-Mismatch}

From source preparation to gate operation to detection,
mode-mismatch is an unavoidable issue in optical quantum computing
that causes unlocated errors which lowers the fidelity of the
device\footnote{Note that mode mismatch in CZ and $\tau$ gate causes
unlocated error that lowers the fidelity, in which we find that it
cannot be improved with state distillation.}. Fortunately, with the
help of quantum error correction, a certain amount of unlocated
error rate, including but not limited to mode-mismatch errors, can
be tolerated. A reliable quantum gate must therefore have unlocated
error rates below this threshold.

The dominant source of mode-mismatch error in the GC-Zeno gate is
from the CZ gate and $\tau$ gate, where two-photon interaction
occurs. Here we follow Rohde et al's~\cite{Rohde06} analysis to
examine the effect of such error. We take the simplest model in
which the mode-mismatch is present between the photons entering the
gate but remain constant through the gate. In this case, the
mode-mismatch in two-photon interaction can be analysed as having
two-photons fail to interact with some probability. This allows us
to write the operations for the CZ gate as follow, where
$0<\Gamma<1$ quantifies the overlap of the two wavepackets.
$\Gamma^2$ is the probability that the two photons successfully
interacted and $\Gamma=0$ for completely mode-mismatched and
$\Gamma=1$ for completely mode-matched. The bar in the
$|\bar{1}1\ra$ term indicates mode-mismatched component of the
state.

\begin{eqnarray}
|00\ra & \rightarrow & |00\ra\nonumber\\
|01\ra & \rightarrow & \sqrt{\gamma_{1}^{n}}|01\ra\nonumber\\
|10\ra & \rightarrow & \sqrt{\gamma_{1}^{n}}|10\ra\nonumber\\
|11\ra & \rightarrow & -\Gamma\gamma_{1}^{n}\tau|11\ra
+\sqrt{1-\Gamma^2}\gamma_{1}^{n}|\bar{1}1\ra\nonumber\\
&&+ f(|02\rangle, |20 \rangle)
\end{eqnarray}

And similarly for the operations of $\tau$ gate:

\begin{eqnarray}
|00\ra & \rightarrow & |00\ra\nonumber\\
|01\ra & \rightarrow & \sqrt{\gamma_{1}^{'}}|01\ra\nonumber\\
|10\ra & \rightarrow & \sqrt{\gamma_{1}^{'}}|10\ra\nonumber\\
|11\ra & \rightarrow & \Gamma\gamma_{1}^{'}\tau|11\ra
+\sqrt{1-\Gamma^2}\gamma_{1}^{'}|\bar{1}1\ra\nonumber\\
&& + f(|02 \rangle, |20 \rangle)
\end{eqnarray}

With the equations for the CZ and $\tau$ gate\footnote{Due to
mode-mismatch, the $\tau$ gate is less effective in two-photon
distillation. It is true that we can increase the two-photon
absorption strength in the $\tau$ gate to make up for the
inefficiency. However, here we assume that we do not know the
mode-matching parameter $\Gamma$ of the input wavepackets, and
therefore this adjustment cannot be made. In addition, increasing
the two-photon distillation will increase single-photon loss as
well, which lowers the probability of success.}, and given a
normalized input state
$(\alpha|00\ra+\beta|01\ra+\delta|10\ra+\epsilon|11\ra)$, we can
obtain analytical expression for the fidelity $F$ (per qubit) and
success probability $P_s$ (per qubit) of the GC-Zeno gate as follow.
Equation~\ref{eqn:F} and~\ref{eqn:Ps} show that both the fidelity
and success probability are state dependent due to mode-mismatch.
The worst case of fidelity occurs when the input state is the equal
superposition state $(|00\ra+|01\ra+|10\ra+|11\ra)/2$ (i.e.
$\alpha=\delta=\beta=\epsilon=1/2$) and the worst case of success
probability occurs when the input state is the pure state $|11\ra$
(i.e. $\alpha=\beta=\delta=0$ and $\epsilon=1$).

\begin{equation}
F=\frac{\alpha^*A_1+\beta^*A_2+\delta^*A_3
+\epsilon^*A_4}{\sqrt{|A_1|^2 + |A_2|^2 +|A_3|^2+|A_4|^2}}
\label{eqn:F}
\end{equation}

\begin{eqnarray}
P_s&=&\frac{e^{-2\lambda/\kappa}\tau^{2/\kappa}}{2(1+e^{-\lambda/\kappa}\tau^{2+1/\kappa})}\nonumber\\
&\times& \sqrt{|A_1|^2 + |A_2|^2 +|A_3|^2+|A_4|^2}\label{eqn:Ps}
\end{eqnarray}

where $a_1=(\tau+\tau\Gamma+\sqrt{1-\Gamma^2})$,
$a_2=(\tau-\tau\Gamma+\sqrt{1-\Gamma^2})$,
$a_3=(\tau-\tau\Gamma-\sqrt{1-\Gamma^2})$,
$a_4=(\tau+\tau\Gamma-\sqrt{1-\Gamma^2})$, and $A_1=\alpha
a_1^2+\beta a_1 a_2 +\delta a_1 a_2 +\epsilon a_2^2$, $A_2=\alpha
a_1 a_3+\beta a_2 a_3 +\delta a_1 a_4 + \epsilon a_2 a_4$,
$A_3=\alpha a_1 a_3 + \beta a_1 a_4 + \delta a_2 a_3 + \epsilon a_2
a_4$, $A_4=\alpha a_3^2+\beta a_3 a_4 +\delta a_3 a_4 +\epsilon
a_4^2$

\subsection{IIb. Effect of Detector Efficiency}

In practice, even for the most advanced photon detector, detector
inefficiency is always present. The effect of this noise is to
reduce the probability of success of the gate but not the fidelity
because the errors are locatable.

% The unlocated error is 1-F
%where the fidelity includes unlocated measurement error of (1-F)/10
%to conform with Dawson et al'a analysis. Similarly, the located
%error is 1-P, where the success probability includes located
%detector failure of (1-P)/10.

%\begin{itemize}
%\item{Amount of mode-mismatch error tolerable with kappa $10^4$}
%\item{The lowest detector efficiency required. Include
%detector efficiency and find the amount of mode-mismatch tolerable
%with kappa $10^4$?}
%
%\end{itemize}

\section{III. Estimate of bounds for Fault Tolerance}
We now wish to estimate lower bounds on the mode-matching, $\Gamma$,
and photon loss ratio, $\kappa$, that will still allow
fault-tolerant operation. We allow a small amount of detector
inefficiency but assume all other parameters are ideal. To make this
estimate we directly use the bounds obtained by Dawson et
al~\cite{Dawson06} for a deterministic error correction protocol.
For this protocol, they numerically derived one bound using the
7-qubit Steane code and another bound using the 23-qubit Golay code.

%By analyzing the two Bell measurement circuits individually with
%Bell state inputs, we find that the circuits give pauli Y and Z
%errors more probable than X errors. The unlikelihood of one type of
%error suggests that the error bounds provided by the deterministic
%protocol is appropriate, if we assume that this property is not
%changed by the entangled resource and the single qubit operations
%after measurements in the GC-Zeno gate.

%Since the GC-Zeno gate measures the input qubits and two of the
%entangled qubits, and does only X and Z corrections on the remaining
%two qubits, therefore the kind of errors that the GC-Zeno gate
%generates are only located and unlocated X and Z errors. These types
%of errors are included in the deterministic error correction
%protocol and

In order to use the Dawson et al's bounds we need to identify the
unlocated and located error rates for our gate. In general, the
unlocated error rate is less than $1-F$ but here we take it to be
$1-F$ because in our analysis, $\gamma$ is almost 1, which means the
other terms involved are very small. The located error rate is
simply $1-P_s$ (both $F$ and $P_s$ are per qubit). Using these
relationships, we convert each of the bounds into a fidelity versus
success probability bound. For a gate built with two-photon
absorbers that have a certain single-photon to two-photon
transmission ratio $\kappa$, we can find an optimal $\lambda$ (i.e.
choosing an optimal absorber length) that gives a maximum success
probability. Hence, by matching the success probability with the
bound, we can determine the corresponding fidelity threshold and
therefore find the least amount of mode-matching required for fault
tolerant gate operation. We note that the error model used by Dawson
et al is specific to their optical cluster state architecture and
will differ in detail from the appropriate error model for the
GC-Zeno gate. Nonetheless we assume that a comparison based on the
total error rates will give a good estimate of the bounds.

\begin{figure}
\centerline{\psfig{figure=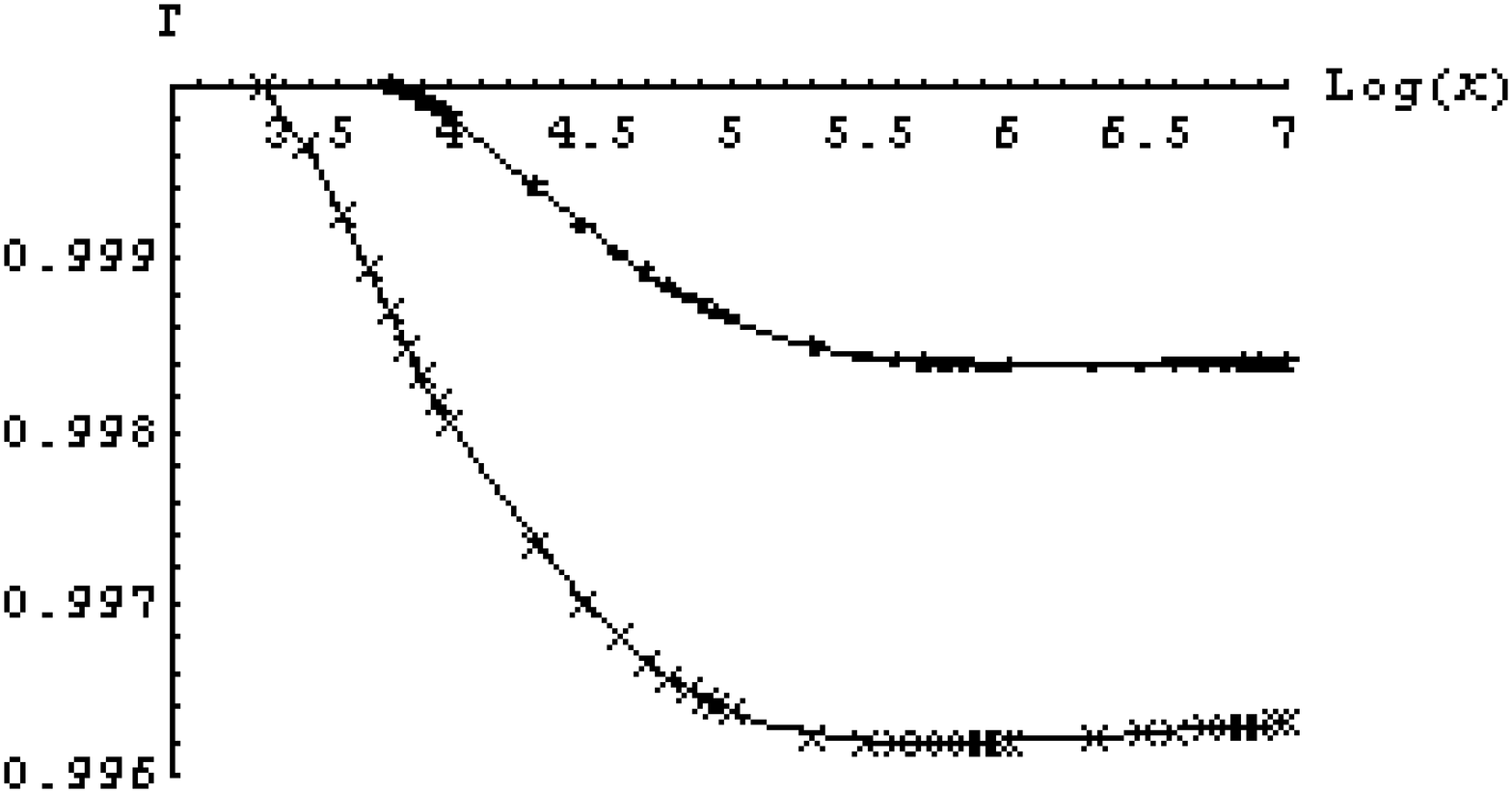,width=3.5cm,angle=90}}
\caption{Lower bounds of amount of mode-matching $\Gamma$ required
for a fault tolerant GC-Zeno gate versus single-photon to two-photon
transmission ratio $\kappa$. The bounds are derived from Dawson et
al's~\cite{Dawson06} results on deterministic error correction
protocol. The top and bottom curves are for the 7-qubit Steane code
and the 23-qubit Golay code respectively. Above the curves are the
regions where the amount of mode-mismatch is tolerable. Here we have
used the worst case input.\\} \label{fig:GammaVsKappaAvg}
\end{figure}

Figure~\ref{fig:GammaVsKappaAvg} shows the lower bounds on the
mode-matching parameter $\Gamma$ for a gate with a certain $\kappa$.
Since the fidelity and success probability are state dependent due
to mode-mismatch, in that figure, we have plotted for the case of
worst fidelity input state (i.e. the equal superposition state). The
top and bottom curves are best fit curves for using the 7-qubit
Steane code and the 23-qubit Golay code respectively. The curves
show that highly mode-matched photons are essential for robust gate
operation. With the worst fidelity input state,
$(|00\ra+|01\ra+|10\ra+|11\ra)/2$, for the Steane code, the lowest
$\Gamma$ required for fault tolerant operation is about 0.998, where
$\kappa=10^6$, and for the Golay code, the lowest $\Gamma$ required
is about 0.996, where $\kappa=5\times10^5$. With the worst success
probability input state, $|11\ra$, for the Steane code, the lowest
$\Gamma$ required for fault tolerant operation is about 0.995, where
$\kappa=10^6$, and for the Golay code, the lowest $\Gamma$ required
is about 0.989, where $\kappa=5\times10^5$.
Figure~\ref{fig:GammaVsKappaAvg} also shows that under (near)
perfect mode-matching, the required $\kappa$ can be as low as
approximately 6000 for the Steane code and 2000 for the Golay code.
Two-photon absorbers with such $\kappa$ values may be achievable
with the best of current technology.\\

\subsection{IIIa. Advantage of Using State Distillation}

State distillation allows us to trade off some success probability
against fidelity for the GC-Zeno gate, or in other words, reducing
the unlocated error rate by having a larger located error rate.
Since the deterministic error correction protocol can tolerate both
unlocated and located errors, therefore we should ask whether state
distillation is truly advantageous? We can answer this question by
comparing two GC-Zeno gates in the case of perfect mode matching,
where one has complete distillation and the other has no
distillation. For the case of complete distillation, the
deterministic error correction protocol with the 7-qubit Steane code
can tolerate errors of a GC-Zeno gate with $\kappa=6100$, and with
the 23-qubit Golay code, it can tolerate errors of a GC-Zeno gate
with $\kappa=2100$. For state distillation to be advantageous under
the same protocol, these values of $\kappa$ must be smaller than the
values of $\kappa$ for the case of no distillation\footnote{An ideal
Zeno gate requires strong quantum Zeno effect and strong quantum
Zeno effect corresponds to a large $\kappa$ value. However such
large non-linearity is difficult to engineer. Hence it is desirable
to have a gate that works with modest $\kappa$.}.

For the case of no distillation, the fidelity and success
probability of the gate becomes state dependent. In the parameters
space of interest, the input that gives the worst fidelity is
$(|00\ra+|01\ra+|10\ra-|11\ra)/2$. With this input state, we find
that for the protocol using the 7-qubit Steane code and no
distillation, the critical $\kappa$ is 12000. Similarly, for the
protocol using the 23-qubit Golay code and no distillation, the
critical $\kappa$ is 4300. With an arbitrary amount of distillation,
the value of $\kappa$ lies between the limit of no distillation and
full distillation cases. Hence it is evident that state distillation
is advantageous. Also, it should be noted it is better to have only
located error, which is the case when there is full distillation,
than have both located and unlocated errors, which is the case when
no or some distillation is utilized.

\section{IV. Conclusion}
In this paper, we have shown that it is possible to build a high
fidelity Zeno CNOT gate with two distillated Zeno gates implemented
in the Gottesman-Chuang teleportation CNOT scheme. For one-photon to
two-photon transmission ratio $\kappa=10^4$ (current best estimate),
the gate has a success probability of 0.76 under perfect
mode-matching. When including measurement noise that equals
one-tenth of the gate's noise, and the effect of mode-mismatch in
the CZ and $\tau$ gate, we find that with the deterministic error
correction protocol using the 7-qubit Steane code, the lowest
$\Gamma$ required for fault tolerant gate operation is 0.998, where
$\kappa=10^6$. For using the 23-qubit Golay code, the lowest
$\Gamma$ required is 0.996, where $\kappa=5\times10^5$. Hence, the
requirement on mode-matching is stringent for a fault tolerant
GC-Zeno gate.

%In this paper, we have modelled Franson et al's CZ gate with a
%succession of $n$ weak beam-splitters followed by two-photon
%absorbers, in the (near) continuous limit of large $n$. We analysed
%this CZ gate for both the ideal two-photon absorption case and the
%incomplete two-photon absorption with single photon loss case,
%giving analytical and numerical results for the fidelity and
%probability of success. The result shows that for a free-standing
%gate we need an absorption ratio $\kappa$ of a million to one to
%achieve $F>0.99$ and 100 million to one to achieve $F>0.999$, where
%recent estimate only suggests that $\kappa\approx10000$ may be
%achievable. We therefore employ this gate for qubit fusion, where
%the requirement for $\kappa$ is less restrictive. With the help of
%partially offline one-photon and two-photon distillations, we can
%achieve a CZ gate with unity fidelity and with probability of
%success is about 0.87 for $\kappa=10000$. We conclude that when
%employed as a fusion gate, the Zeno gate could offer significant
%advantages over linear techniques for reasonable parameters.\\

%\textbf{Acknowledgement}\\
%We thank W.J.Munro, A.Gilchrist and C.Myers for useful discussions.
%This work was supported by the Australian Research Council and the
%DTO-funded U.S. Army Research Office Contract No. W911NF-05-0397.\\

\end{document}